\documentstyle[preprint, epsfig,eqsecnum,aps]{revtex}
\tightenlines

\begin{document}
\draft
\title{Description of Single and Double analog States in the f$_{7/2}$ shell
 - the Ti isotopes}

\author{A. A. Raduta$^{1)}$, L. Zamick$^{2)}$, E. Moya de Guerra$^{3)}$,
$^{4)}$Amand Faessler and P. Sarriguren$^{3)}$ }
\address{$^{1)}$Institute of Physics and Nuclear Engineering, Bucharest POBox MG6, Romania and Theoretical Physics and Mathematics, Faculty of Physics, Bucharest University, Bucharest, POBox MG11, Romania}
\address{$^{2)}$Department of Physics and Astronomy, Rutgers University, Piscataway, New Jersey USA 08855-8019}
\address{$^{3)}$ Instituto de Estructura de la Materia, CSIC, Serrano 119-123, 28006 Madrid, Spain}
\address{$^{4)}$ Intitut fuer Theoretische Physik der Universitaet Tuebingen,Auf der Morgenstelle 14, Tuebingen, Germany}
\maketitle
\date{\today}

\begin{abstract}
The excitation energies of single analog states in even-odd  Ti isotopes and double analog states in even-even Ti isotopes are microscopically described in a single j-shell formalism. A projection procedure for generalized BCS states has been used. As  an alternative description a particle-core formalism is proposed. The later picture provides a two parameter expression for excitation energies which describes fairly well the data in four odd and three even isotopes of Ti.\end{abstract}
\pacs{PACS numbers: 21.10.-k, 27.40.+z, 21.60.Cs}

\section{Introduction}
\label{sec:level1}
In a series of papers [1-5], Zamick and his collaborators investigated the
spectra of even-even and even-odd nuclei in the pf region. The authors used
the shell model calculations in a single open shell for both protons and
neutrons. They remarked that for some even-even nuclei, the $J^{\pi}=0^+$
excited states  were at
twice the energies of $J=j$ excited states in the neighboring even-odd nuclei,
in a single j shell calculation for any isospin conserving interaction.
Examples of this type are the pairs ($^{44}$Ti,$^{43}$Sc), ($^{44}$Ti,$^{43}$Ti),
($^{48}$Ti,$^{49}$Ti), ($^{48}$Ti,$^{47}$Sc). In  order that this relationship
 holds it is necessary that
the diagonalization spaces for the odd and even systems are equal to each other.
 For the nuclei mentioned above the prove was analytically given in ref.
 \cite{ZaDe}. The states mentioned above for the even-odd and even-even isotopes can be viewed as single (SA) and  double analog (DA) states, respectively.

For the combination ($^{46}$Ti, $^{45}$Sc) and the cross conjugate pair($^{50}$Cr, $^{51}$Cr) one almost gets a 2 to 1 relation, which becomes precisely 2 to 1 if one does not allow admixtures of seniority $v=4$ states in the ground states of the even-even nuclei. In ref. \cite{Devi} the excitation energies of 
higher isospin states were parametrised by linear relations
\begin{eqnarray}
E^*_{T+1}&=&b(T+X),~~{\rm For} ~J=j,~{\rm SA} (\rm{ Single~ Analogue~ State}),\nonumber\\
E^*_{T+2}&=&2b(T+X+\frac{1}{2}), ~~{\rm For} ~J=0,~ {\rm DA} (\rm {Double~ Analogue~ State}).
\end{eqnarray}
These are equivalent to a quadratic expression for the binding energies
$BE=-2bT(T+Y)$, where $Y=2X-1$. For Wigner SU(4) model Y is equal to 4 and X=2.5.

 In the present paper we revisit the above problem for the isotopes of Ti.
 We suppose that the states of interest are mainly determined by the valence
 nucleons in the $f_{7/2}$ shell, which interact among themselves through an
 isospin invariant pairing force.

 The SA states are populated in a (p,n) reaction process\cite{AHK,Ring}.
 Thus, the ground state ($T,T_3=T$) of the target nucleus  ($N,Z$)
 (often called  mother nucleus), is transformed by the isospin raising operator \footnote{Throughout this paper we consider that protons and neutrons have $T_3$ equal to $\frac{1}{2}, -\frac{1}{2}$ respectively.}
  $T_+$ to the
 state ($T,T-1$) describing the daughter nucleus ($N-1,Z+1$). This state is
 certainly an
 excited state, the daughter nucleus ground state  being  a ($T-1,T-1$) state.
 The SA state has the same quantum numbers as the ground state of the
 mother nucleus except for $T_3$ which is by one unit smaller than the initial
 value. This state can be described either by a neutron hole-proton particle
 Tamm-Dancoff approximation (TDA) applied to the mother system or by
 diagonalizing a
 many body Hamiltonian associated to the daughter nucleus within a
 $|\alpha TT_3\rangle$ basis. Considering now the daughter nucleus in the
 SA state as a
 target, one may define again, by a (p,n) process, a new SA state. This is a
 DA state
 with respect to the initial mother nucleus and has the isospin quantum numbers
 (T,T-2). According to the above considerations this state can be obtained by
 two successive TDA phonon excitations of the ground state of the mother
 nucleus.
 Roughly speaking the SA energy is about the shift of proton single
 particle energy due to the Coulomb interaction and symmetry energy, the
 deviation being caused by the proton-neutron residual interaction.
 The remark of Zamick and his collaborators is that in some Ti even isotopes
 the excitation energy of the DA states is twice as much as the excitation energy
 of SA in the neighboring even-odd Ti isotopes. For illustration, an example
 is given in Figs. 1, 2. The question is of course, whether this is a general
 feature or is mainly determined by the interaction regime (strong or weak) between
 the odd nucleons and the core on one side and between odd nucleons on
 other side.

The states whose energies are suspected to provide a ratio of two to one
are considered as approximate eigenstates of the model Hamiltonian,
which will be treated within a generalized BCS formalism.
The corresponding energies are obtained as average values of the many body
Hamiltonian on  $N,T,T_3$ projected states, generated by the generalized
BCS ground state. As in a previous publication \cite{ApEl}, we restrict the space of
single particle states to a proton single $j_p$ and a neutron single $j_n$
shell with $j_p=j_n$.

The results are presented as follows. In Section II, the even-even system is
presented.
Since the results for such systems were earlier described in ref. \cite{ApEl},
here we
collect only the main results in order to facilitate a self-consistent
picture.
Section III is devoted to the even-odd system. Numerical results are commented
in Section IV and the final conclusions are given in Section V.
\section{Even-even system}
\label{sec:level2}
The many body systems for mother and daughter nuclei are described by the following
Hamiltonian:
\begin{equation}
H=\epsilon(\hat{N}_p+\hat{N}_n)-\frac{G}{4}\sum_{j,j^\prime}
(P^{\dagger}_p(j)P_p(j^{\prime})+P^{\dagger}_n(j)P_n(j^{\prime})+
2P^{\dagger}_{pn}(j)P_{pn}(j^{\prime})).
\end{equation}
where the following notations have been used:
\begin{eqnarray}
P^{\dagger}_{\tau}(j)&=&\sum_{m}c^{\dagger}_{\tau jm}c^{\dagger}_{\widetilde{\tau jm}},
\nonumber\\
P^{\dagger}_{pn}(j)&=&\sum_{m}c^{\dagger}_{pjm}c^{\dagger}_{\widetilde{njm}}.
\end{eqnarray}
In a previous paper, two of us (A.A.R. and E.M.G.) approached the generalized BCS
ground state by a wave function obtained by rotating the ground state of a
proton-neutron pairing Hamiltonian in the isospin space. Rotation angles
depend on the quantum numbers defining the proton-neutron pair.
\begin{equation}
|BCS\rangle=\prod_{jm}\hat{R}(\Omega^{(j)}_0)(U_j+V^*_jC^{\dagger}_{pjm}
c^{\dagger}_{\widetilde{njm}})|0\rangle \equiv \hat{R}(\Omega_0)|BCS\rangle_{pn},
\end{equation}

\noindent
where the standard notation for the BCS ground state of the p-n pairing
Hamiltonian,
$|BCS\rangle_{pn}$, and the rotation in the space of the many body states:
\begin{equation}
\hat{R}(\Omega_0) = \otimes_{j}\hat{R}(\Omega^{(j)}_0),
\end{equation}
have been used.
The transformation $\hat{R}(\Omega_0)$ has the meaning of a collective rotation which
fixes the orientation of the intrinsic-like frame in the space of isospin.
Thus $|BCS\rangle$ plays the role of an intrinsic state while $|BCS\rangle_{pn}$
of an auxiliary intrinsic state.
Clearly the function $|BCS\rangle$ breaks the gauge and isospin symmetries.
These symmetries can be restored by a projection procedure, and the resulting
states are:
\begin{equation}
|NTMK\rangle=\frac{2T+1}{16\pi^3}{\cal N}_{NT}
\int {D^T_{MK}}^*e^{i(\hat{N}-N)\Phi}\hat{R}(\Omega)\hat{R}(\Omega_0)
|BCS\rangle_{pn}d\Omega d\Phi.
\end{equation}
Here N stands for the total number of valence nucleons, T is the total isospin,
K the third component of the total isospin (equal to minus the half of the neutron excess). 
As we already mentioned we suppose that the excitation energies are determined by
the valence nucleons. In the case of Ti isotopes, the open shells for protons
and neutrons are $f_{\frac{7}{2}}$.
In what follows we give, therefore, the useful results for the case of a single
j-shell. To simplify the notations we omit the index j specifying the shell.
Semi-degeneracy of the j-shell is denoted by $\Omega=\frac{2j+1}{2}$

The norm for the state $|NTMK\rangle $  is:
\begin{equation}
{\cal N}^{-2}_{NT}=|D^T_{K,0}(\Omega_0)|^2(N_{pn})^2(\frac{|V|}{U})^N\sum_p
(-2)^p\left( \matrix{\Omega\cr p}\right)\left(\matrix{2\Omega-2p \cr
\frac{N}{2}-2p}\right)Z(\frac{N}{2},p).
\end{equation}
The projected states energy characterizing the even-even system of valence nucleons
is analytically given by:

\begin{equation}
\langle NTMK|H|NTMK\rangle = \epsilon N-G\frac{N}{4}\left(2\Omega-
\frac{N}{2}+1\right)-\frac{G}{2}{\cal N}^2_{NT}N^2_{pn}
|D^T_{K0}(\Omega_0)|^2\sum_p S_pZ(\frac{N}{2},p).
\end{equation}
with the  notations:
\begin{eqnarray}
& &S_p  = \left(\frac{|V|}{U}\right)^N
\\
&\times&\left[4p \left(\Omega  -  \frac{N}{2}+p+1\right )(-2)^p
\left(\matrix{\Omega \cr p}\right ) \left(\matrix{2\Omega -2p
\cr \frac{N}{2}-2p}\right)+4(-2)^p(p+1)^2
\left(\matrix{\Omega \cr p+1}\right ) \left(\matrix{2\Omega -2(p+1)
\cr \frac{N}{2}-2(p+1)}\right)\right]\nonumber
\end{eqnarray}
The function $Z(\frac{N}{2},p)$ is defined in Appendix A
\footnote{In ref.\cite{ApEl} the second term in Eq. (2.8) has, by error, a factor 2 instead of 4. For the angular momentum considered there the deviation from the correct result is however very small.}

\section{Even-odd system}
\label{sec:level3}

\subsection{Extension of projection formalism}

The projected state describing the even-odd system is defined by:
\begin{equation}
|N+1,T_oM_oK_o\rangle={\cal N}_{N+1,T_o}\frac{2T_o+1}{16\pi ^3}
\int e^{i(\hat{N}-(N+1))\phi}{D^{T_o}_{M_oK_o}}^*\hat{R}(\Omega)\hat{R}(\Omega_0)
c^{\dagger}_{njm}|BCS\rangle_{pn}d\Omega d\phi.
\end{equation}
The norm of the state $|N+1T_oM_oK_o\rangle$ has the expression:
\begin{eqnarray}
&&{\cal N}^{-2}_{N+1,T_o}=|D^{To}_{K_o,-\frac{1}{2}}(\Omega_0)|^2
(N_{pn})^2(\frac{|V|}{U})^N\sum_p
(-2)^p\left( \matrix{\Omega -1\cr p}\right)
\nonumber\\
&\times&\left[
\left(\matrix{2\Omega-2p-1 \cr\frac{N}{2}-2p}\right)
Z^{(\nu)}(\frac{N}{2},p)
-\sqrt{2}\left(\matrix{2\Omega-2p-2 \cr\frac{N}{2}-2p-1}\right)
Z^{(\pi)}(\frac{N}{2},p)\right],~N=even.
\end{eqnarray}
where the functions $Z^{(\tau)}(\frac{N}{2},p)$ are those defined in Appendix A.
 Alternatively, the norm of the projected state  describing an odd number of nucleons
can be expressed in terms of norms associated to the even system:
\begin{equation}
{\cal N}^{-2}_{N+1,T_o}=U^2|D^{To}_{K_o,-\frac{1}{2}}(\Omega_0)|^2
\sum_{T_e} {\cal N}^{-2}_{NT_e}\left(C^{T_e \frac{1}{2} T_o}_{0~
\frac{1}{2} ~\frac{1}{2}}\right)^2.
\end{equation}
Following the procedure described in ref.\cite{ApEl}, one can express the energy
 for the even-odd system as:
\begin{eqnarray}
&&\langle N+1T_oM_oK_o|H|N+1T_oM_oK_o\rangle =\nonumber\\
&& \epsilon (N+1)-
\frac{G}{2}\frac{N}{2}\left(2\Omega-
\frac{N}{2}\right)-\frac{G}{4}{\cal N}^2_{N+1T_o}N^2_{pn}\left(\frac{|V|}{U}
\right)^N
|D^{T_o}_{K_o,-\frac{1}{2}}(\Omega_0)|^2
\nonumber\\
&\times&\sum_p\left \{Z^{(\nu)}(\frac{N}{2},p)4(-2)^p
\left[p(2\Omega-N+2p+1)
\left( \matrix{\Omega -1\cr p}\right)
\left(\matrix{2\Omega-2p-1 \cr\frac{N}{2}-2p}\right)\right.\right.
\\
&+&\left. 2(p+1)^2
\left( \matrix{\Omega -1\cr p+1}\right)
\left(\matrix{2\Omega-2(p+1)-1 \cr\frac{N}{2}-2(p+1)}\right)
+(p+1)
\left( \matrix{\Omega -1\cr p}\right)
\left(\matrix{2\Omega-2p-2 \cr\frac{N}{2}-2p-1}\right)\right]
\nonumber\\
&+& Z^{(\pi)}(\frac{N}{2},p)\frac{4(-2)^{p+1}}{\sqrt{2}}
 \left[(2p+1)
\left(\Omega-\frac{N}{2}+p+1\right)
\left( \matrix{\Omega -1\cr p}\right)
\left(\matrix{2\Omega-2p-2 \cr\frac{N}{2}-2p-1}\right)\right.
\nonumber\\
&+&2(p+1)(p+2)
\left. \left.\left( \matrix{\Omega -1\cr p+1}\right)
\left(\matrix{2\Omega-2(p+1)-2 \cr\frac{N}{2}-2(p+1)-1}\right)
-(p+1)
\left( \matrix{\Omega -1\cr p+1}\right)
\left(\matrix{2\Omega-2(p+1)-1 \cr\frac{N}{2}-2(p+1)}\right)\right]\right\}.
\nonumber
\end{eqnarray}
with the  factors Z  defined in Appendix A. Also the notation 
$\Omega$ for the state semi-degeneracy is used.

\subsection{Particle core interaction}

The odd system can be described within a particle-core coupled basis:
\begin{equation}
|N+1T_oM_0K_0\rangle=\sum_{M_e,\mu}C^{T_e~\frac{1}{2}~T_o}_{M_e~ \mu ~M_o}
|NT_eM_eK_e\rangle \chi_{\mu},
\end{equation} 
where $|NT_eM_eK_e\rangle$ describes the even-even system and is given by
Eq. (2.5) while $\chi_{\mu}$ stands for the odd neutron wave function in the laboratory frame. Of course the  isospin projections in the intrinsic frame satisfy the equation
\begin{equation}
K_o=K_e-\frac{1}{2}
\end{equation}
Since the even system states are $J=0$ states, it results that the odd system states are $J=j$ states

In order to treat the many body Hamiltonian (2.1) in the basis defined above it is convenient to separate it into three parts describing the two components of the system as well as their mutual interaction:
\begin{equation}
H=H_{sp}+H_{core}+H_{coup}
\end{equation}
 We assume that the coupling term is simulated by a scalar operator with respect to isospin rotations:
\begin{equation}
H_{coupl}=2F \vec{t}\vec{T},
\end{equation}
where $\vec{t}$ and $\vec{T}$ are the isospin operators acting on the state describing the odd particle and the even-even core, respectively.

As can be seen from Table 1, each state of the odd system can be obtained uniquely by coupling the odd neutron isospin to a single isospin state from the core space. For example the isospin 3/2 of $^{47}$Ti can be obtained by coupling the odd neutron isospin to the state of the even core with T=1 while the state 5/2 originates from the coupling of the odd particle to the $T=3$ core state.
Another remark which is pertinent is that the low isospin state in the odd system is obtained by aligning the odd neutron isospin to the core isospin.
By contrary for the higher isospin state an anti-alignment of the two isospin takes place. This property holds for all even-odd pairs of isotopes considered here. As a consequence the particle core interaction is repulsive in lowest isospin states and attractive in the higher isospin ones.
For particle core coupling models it is customary to adopt the approximations
\begin{eqnarray}
\langle N+1 T_oM_oK_o|H_{core}|N+1T_oM_oK_o\rangle &=&
\langle N T_eM_eK_e|H_{core}|NT_eM_eK_e\rangle ,
\nonumber\\
\langle N+1 T_oM_oK_o|H_{sp}|N+1T_oM_oK_o\rangle &=& \epsilon  
\end{eqnarray}
The average for the core Hamiltonian has been calculated in Section II, being given by the Eq. (2.7).
In this way the excitation energies in odd isotopes are related to those in the neighboring even-even ones by:
\begin{eqnarray}
\Delta E(\frac{7}{2}, \frac{5}{2})=\Delta E(4,2)-7F,
\nonumber\\
\Delta E(\frac{5}{2}, \frac{3}{2})=\Delta E(3,1)-5F,
\nonumber\\
\Delta E(\frac{3}{2}, \frac{1}{2})=\Delta E(2,0)-3F.
\end{eqnarray}

\section{Numerical results}
Here we consider the SA states in odd  and DA states in even mass isotopes of Ti.The energies for the lowest isospin states in each case are obtained as expectation values of the many body Hamiltonian on the particle number and isospin projected states. Excitation energies  depend on the pairing strength G. This is fixed so that the experimental values of the excitation energies are 
reproduced. In this way we want to explore effects like the dependence of 
G on the atomic mass number, the blocking due to the odd particle as well as whether the v=0 (v denotes seniority quantum number) description suffices for both even-even and even-odd nuclei. As we stated from the beginning we restrict 
the space of single particle states to a single j shell, i.e. $f_{\frac{7}{2}}$. The excitation energies obtained in this way are given in Table 1.
The values of G obtained through the fitting procedure mentioned above are plotted in Figs. 3-6. as function of A. In the first figures we omitted the value of G for $^{45}$Ti which is quite large ($\approx 8.35$MeV).  The large values of G for the lowest mass odd isotopes might be caused by the fact that the restriction of the many body states to the v=0 subspace is unrealistic and the neglected contribution is compensated through an artificial increase of the paring strength.
Also the blocking effect of the odd particle which,  as a matter of fact, is neglected here, is suspected to be large. For the last two pairs of isotopes ($^{46}$Ti,$^{47}$Ti) and ($^{48}$Ti, $^{49}$Ti) the pairing strengths are fairly close to each other. Moreover the ordering of the G value for even-even and even-odd isotopes is opposite to that of the case of light isotopes. As shown in Fig. 4 the pairing strength falls on a third degree polynomial in A. The  A dependence of 
G is plotted separately for even-even and even-odd isotopes in Figs. 5 and 6 respectively. While in the even A case the points are almost collinear, for odd isotopes the pairing strengths lie on a Gaussian curve.

\begin{flushleft}
\begin{tabular}{ccccc}\hline
        &   &   &Single j shell&Experiment\\
Even-Odd&$\hskip1cm$   & $\hskip1cm$  &E(MeV)  &  E(MeV) \\
$^{43}$Ti&  1/2&  3/2& 4.340 & 4.338\\
$^{45}$Ti&  1/2&  3/2& 4.178 & 4.176\\
         &     &     & 4.725 &      \\
$^{47}$Ti&  3/2&  5/2& 7.191 & 7.187\\
$^{49}$Ti&  5/2&  7/2& 8.725 & 8.724\\
Even-Even&     &     &       &       \\
$^{44}$Ti&   0 &   2 & 9.345 & 9.340 \\
$^{46}$Ti&   1 &   3 & 14.150& 14.153\\
$^{48}$Ti&   2 &   4 & 17.395& 17.398\\
\hline
\end{tabular}
\end{flushleft}
{\small {Table 1.
Excitation energies of single and double analog states in Ti isotopes (MeV)}}

It is worth mentioning that the isospin invariant pairing Hamiltonian 
($H_{pair}$) considered for a single j-shell is exactly solvable. Indeed, $H_{pair}$ can be expressed in terms of the quadratic Casimir operator of the group O(5) generated by the proton-proton, neutron-neutron and proton-neutron quasispin operators 
\cite{Taka}. Therefore, its eigenvalues have simple expressions 
in terms of the highest weight of the irreducible representations of the 
O(5) group. These weights are determined by the reduced isospin t 
\cite{Flow} and the seniority quantum number v.

\begin{equation}
H_{pair}=-\frac{G}{2}\left (\frac{1}{4}(N-v)(4\Omega -N-v +6)+t(t+1)-T(T+1)\right).
\end{equation}
The notation $\Omega$ stands for the state semi-degeneracy.
In the BCS treatment adopted here the projected states have vanishing
seniority and reduced isospin.
Therefore for the even-even isotopes one obtains a close formula expressing the energy of a system of N valence nucleons:
\begin{equation}
E_T=\epsilon N-\frac{G}{2}\left(\frac{N}{4}(4\Omega-N+6)-T(T+1)\right)
\end{equation}
Using this formula with G values obtained in the manner specified above one obtains for
excitation energies exactly the same energies as given in Table 1. One may conclude that the projection procedure yields for single j-energy the same value as the group theory method.

Now let us turn our attention to the particle-core formalism.
We applied Eq. (18) for odd isotopes in the following way. The strength of the particle-core interaction was fixed so that the excitation energies for the SA states are reproduced. The values of F, obtained in this way, are plotted in Fig. 7. They are also listed in Table 2. From Fig. 7 one sees that apart from F of $^{45}$Ti the other three points lye on a straight line. However the
 point for $^{45}$Ti is not far from the straight line. Indeed, taking   
the value of F which falls on the straight line (this value is equal to
1.54 MeV and is given in table 2 in brackets)  and calculating the corresponding excitation energy for the state with $T=\frac{3}{2}$, one obtains  a value equal to 4.725 MeV which has to be compared with the experimental value of 4.176
MeV. 
\begin{flushleft}
\begin{tabular}{cccc}\hline
Even-Even&    G (MeV)   &Even-Odd  &F(MeV) \\
$^{44}$Ti&  3.115       &$^{43}$Ti  & 1.669\\
         &              &$^{45}$Ti  & 1.723 (1.54)\\ 
$^{46}$Ti&   2.830      & $^{47}$Ti & 1.392\\
$^{48}$Ti&   2.485      & $^{49}$Ti & 1.239\\
\hline
\end{tabular}
\end{flushleft}

{\small {Table 2.
The values of pairing (second column) and particle-core strengths(fourth column) are given in units of MeV}}

Comparing the values of F and G given in Table 2 one remarks that F is almost
half the value of G.

Coming back to Eq. (18) and using $F=\frac{G}{2}$ one obtains for the excitation energies in the odd isotopes the values 4.672, 4.672, 7.104 and 7.698 MeV , respectively. We remark that the last two  values listed above are close to the experimental values in $^{47}$Ti and $^{49}$Ti, respectively, while for $^{43}$Ti and $^{45}$Ti the deviations are equal to 334 and 496 keV respectively.
Thus one could state that we obtained a quite reasonable description of 
the excitation energies of SA and DA states in Ti isotopes using only one parameter, the pairing strength, which is depending linearly on A.

Denoting by T=$\frac{|N-Z|}{2}$ the ground state isospin, one arrives at very simple expressions for SA and DA excitation energies in terms of isospin:
\begin{eqnarray}
E_{DA} &=& G(2T+3),
\nonumber\\
E_{SA} &=& G(T+1),
\nonumber\\
G      &=&-0.156A+9.979 [MeV], ~~\rm{A=even}
\end{eqnarray}
For even-odd isotopes one should take the expression of G the value of A corresponding to the neighboring even-even-core.
If the small deviation of F from $\frac{G}{2}$ is not ignored then the SA excitation energy is given by:

\begin{eqnarray}
S_{SA} &=& 2(G-F)(T+1),
\nonumber\\
F &=& -0.0717A+1.9773 [MeV],~~\rm{ A=odd}.
\end{eqnarray}

Recalling that $A=2Z\pm 2T$ where the sign minus holds for $^{43}$Ti, it is clear
that the above equations provide quadratic expressions of T for SA and DA excitation energies. 
\section{Conclusions}

Excitation energies for SA (Single Analog) and DA (Double Analog) states in even-odd and even-even isotopes of Ti, respectively, have been described with a many body isospin invariant
Hamiltonian in a single j-shell. The states describing the neighboring even-even and even-odd isotopes are obtained through an N, T projection procedure from a generalized proton-neutron BCS wave function. The pairing strength for even-even systems depend linearly on the atomic mass number A. It is worth mentioning that the group theory formula produces identical results for excitation energies as the projection procedure.
Alternatively the even-odd systems are described in a particle-core coupling scheme where the coupling interaction is an isospin invariant operator acting on both odd particle and core degrees of freedom. The fitted strength for the coupling term is almost half the pairing strength. Taking into account that the
excitation energy for the even system is analytically given by a very simple formula, it results that the present paper provides also a compact expression for the excitation energy in even-odd isotopes. Since G depends linearly on A one may conclude that the excitation energies in the seven isotopes considered here, are fairly well obtained by fixing only two free parameters. Finally, analytical expressions for a quadratic dependence on T for SA and DA excitation energies are obtained.
\section{Appendix A}
Here we give the analytical expressions for the overlap integrals which enter the
formulae determining the energies for even-even and even-odd systems.
\begin{eqnarray}
&&Z\left(\frac{N}{2},p\right)\equiv\frac{2T+1}{2}\int d^T_{00}
\left(d^1_{00}\right)^{\frac{N}{2}-2p}\left(d^1_{10}d^1_{-10}\right)^p
\sin{\beta}d\beta
\nonumber \\
&=&\sum_{k,m,n} (-)^{p+k+n}\frac{(2T+1)2^{-T-p}}{m+n+s+1}\left(\matrix{
T \cr k}\right)^2\left(\matrix{T+p-k \cr m} \right )
\left(\matrix{k+p \cr n} \right).
\nonumber \\
& &m+n+s=even;~~s=\frac{N}{2}-2p
\end{eqnarray}
\begin{eqnarray}
&&Z^{(\nu)}\left(\frac{N}{2},p\right)\equiv\frac{2T_o+1}{2}\int
d^{T_o}_{-\frac{1}{2},-\frac{1}{2}}
\left(d^1_{00}\right)^{\frac{N}{2}-2p}\left(d^1_{10}d^1_{-10}\right)^p
d^{\frac{1}{2}}_{-\frac{1}{2},-\frac{1}{2}}\sin{\beta}d\beta
\nonumber\\
&=&\sum_{k,m,l} (-)^{p+k+m}\frac{(2T_o+1)2^{\frac{N}{2}-T_o-\frac{5}{2}p
+l+m+1}}{\frac{N}{2}-2p+l+m+1}\left(\matrix{
T_o+\frac{1}{2} \cr k}\right)
\left(\matrix{
T_o-\frac{1}{2} \cr k}\right)
\left(\matrix{T_o+p-k+\frac{1}{2} \cr l} \right )
\left(\matrix{k+p \cr m} \right),
\nonumber \\
& &\frac{N}{2}+l+m=even;
\nonumber \\
&&Z^{(\pi)}\left(\frac{N}{2},p\right)\equiv\frac{2T_o+1}{2}\int
d^{T_o}_{-\frac{1}{2},-\frac{1}{2}}
\left(d^1_{00}\right)^{\frac{N}{2}-2p-1}\left(d^1_{10}d^1_{-10}\right)^p
d^1_{-10}d^{\frac{1}{2}}_{\frac{1}{2},-\frac{1}{2}}\sin{\beta}d\beta
\nonumber \\
&=&\sum_{k,m,l} (-)^{p+k+m}\frac{(2T_o+1)2^{\frac{N}{2}-T_o-\frac{5}{2}p
+l+m-1}}{\frac{N}{2}-2p+l+m}\left(\matrix{
T_o+\frac{1}{2} \cr k}\right)
\left(\matrix{
T_o-\frac{1}{2} \cr k}\right)
\left(\matrix{T_o+p-k+\frac{1}{2} \cr l} \right )
\left(\matrix{k+p+1 \cr m} \right).
\nonumber \\
& &\frac{N}{2}+l+m=odd.
\end{eqnarray}

\begin{figure}[h]
\centerline{\psfig{figure=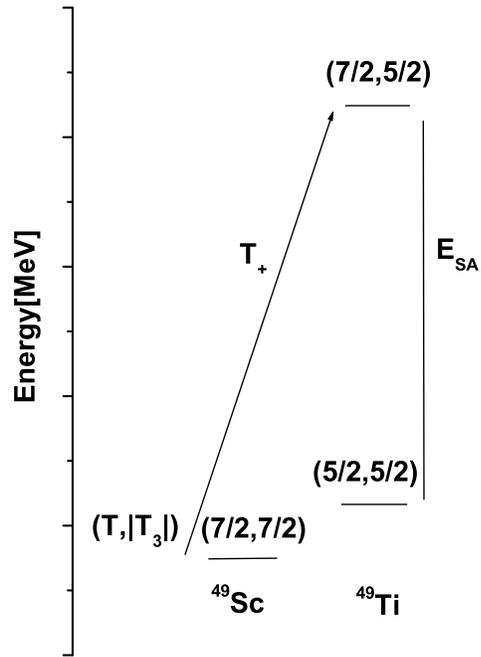,width=10cm,bbllx=2cm,%
bblly=5cm,bburx=20cm,bbury=26cm,angle=0}}
\vspace*{-1cm}
\caption{\label{Fig. 1}The SA state in the daughter nucleus $^{49}$Ti
and the ground state of the mother nucleus $^{49}Sc$ are schematically
presented.}
 
\end{figure}

\newpage

\begin{figure}[h]
\centerline{\psfig{figure=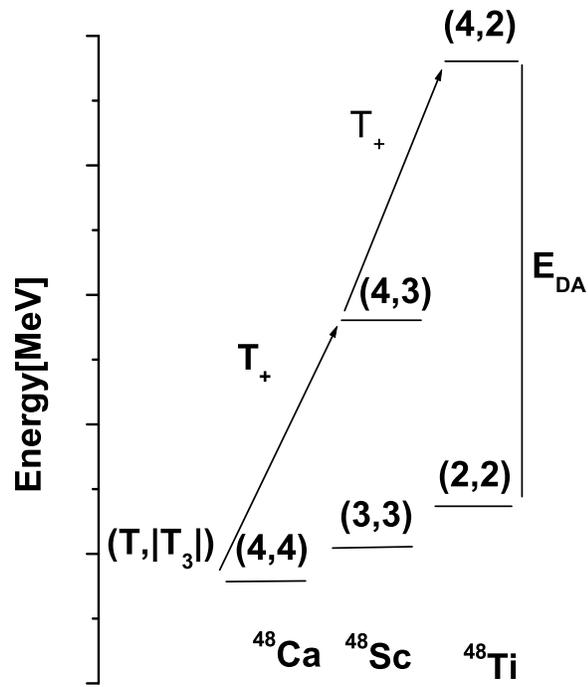,width=10cm, bbllx=2cm%
bblly=5cm,bburx=20cm,bbury=26cm,angle=0}}
\vspace*{-2cm}
\caption{\label{Fig. 2}The DA state in the final nucleus $^{48}$Ti
and the ground state of the initial nucleus $^{48}Ca$ are schematically
presented.}
\end{figure}

\newpage

\begin{figure}[h]
\centerline{\psfig{figure=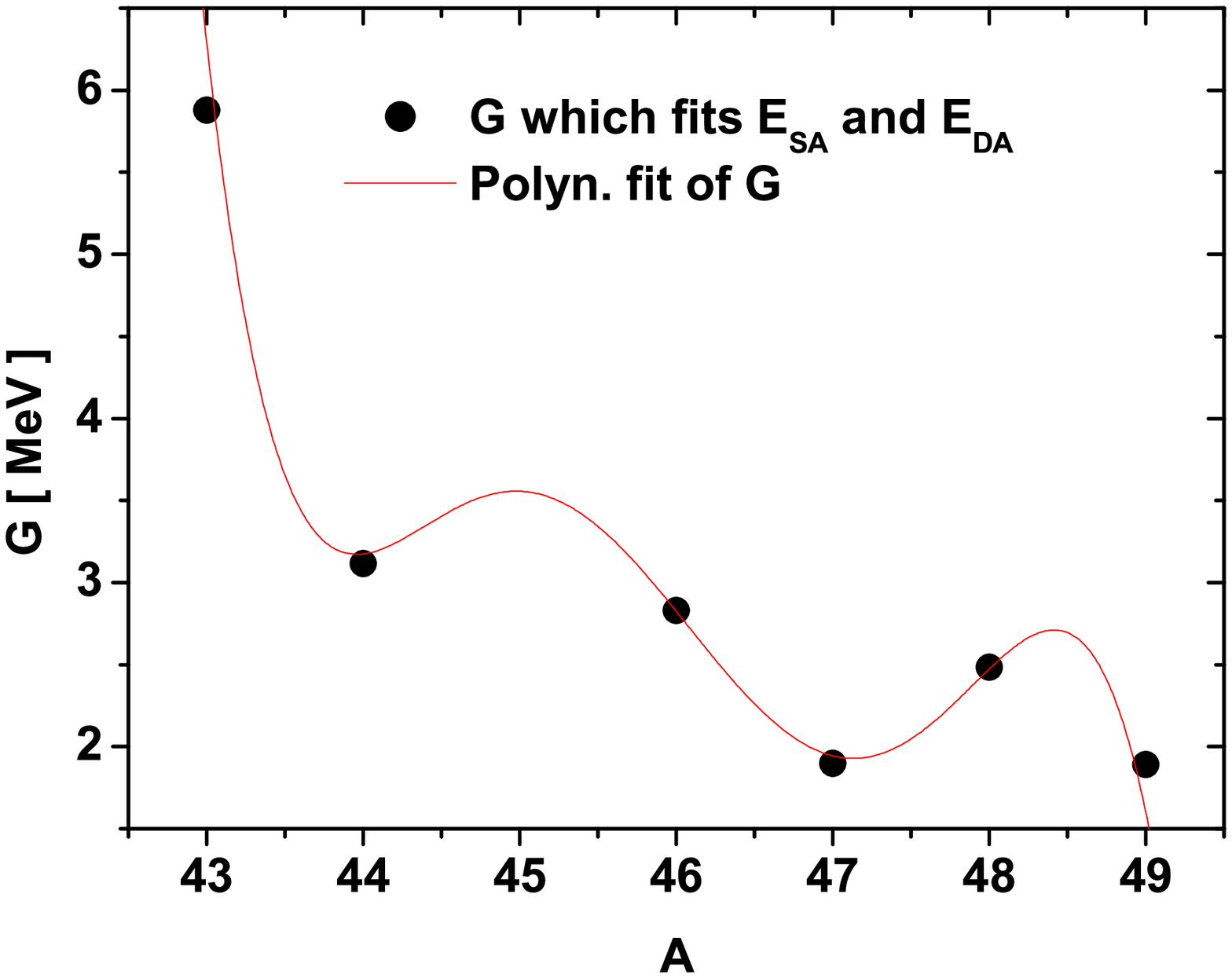, width=14cm, bbllx=2cm%
,bblly=5cm,bburx=20cm,bbury=26cm,angle=0}}
\vspace*{-3cm}
\caption{\label{Fig. 3}(Color online) The pairing strength as function of the atomic mass A.
The values of G are obtained by fitting the single and double analog resonances respectively,
and represented by black circles. These points may be interpolated by a polynomial of
fifth order in A. The resulting polynomial is given by the full line.}
\end{figure}

\newpage

\begin{figure}[h]
\centerline{\psfig{figure=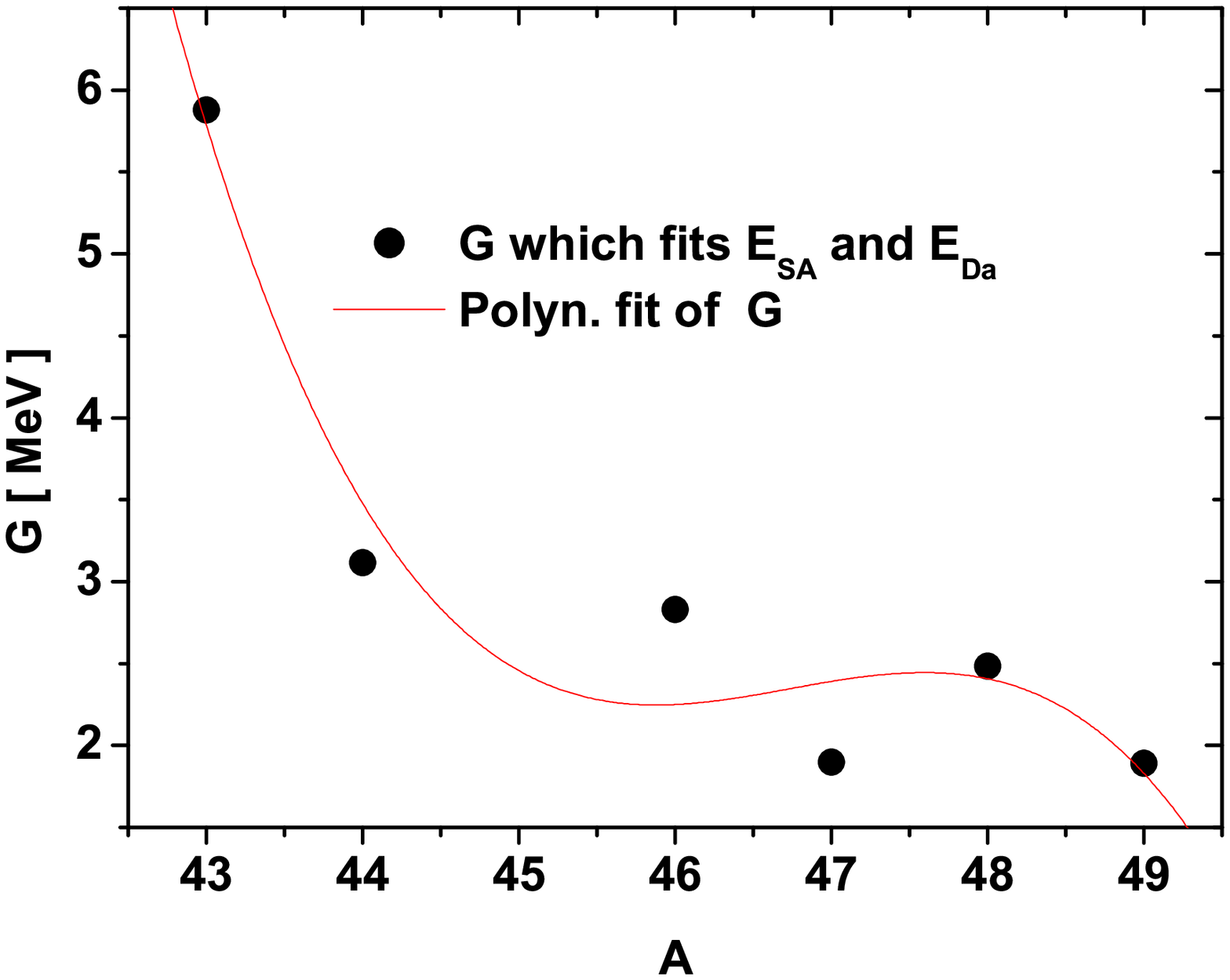, width=14cm, bbllx=2cm%
,bblly=5cm,bburx=20cm,bbury=26cm,angle=0}}
\vspace*{-3cm}
\caption{\label{Fig. 4}(Color online) The pairing strength as function of the atomic mass A.
The values of G are obtained by fitting the single and double analog resonances respectively,
and represented by black circles. These points may be interpolated by a polynomial of
third order in A. The resulting polynomial is given by the full line.}
\end{figure}

\newpage

\begin{figure}[h]
\centerline{\psfig{figure=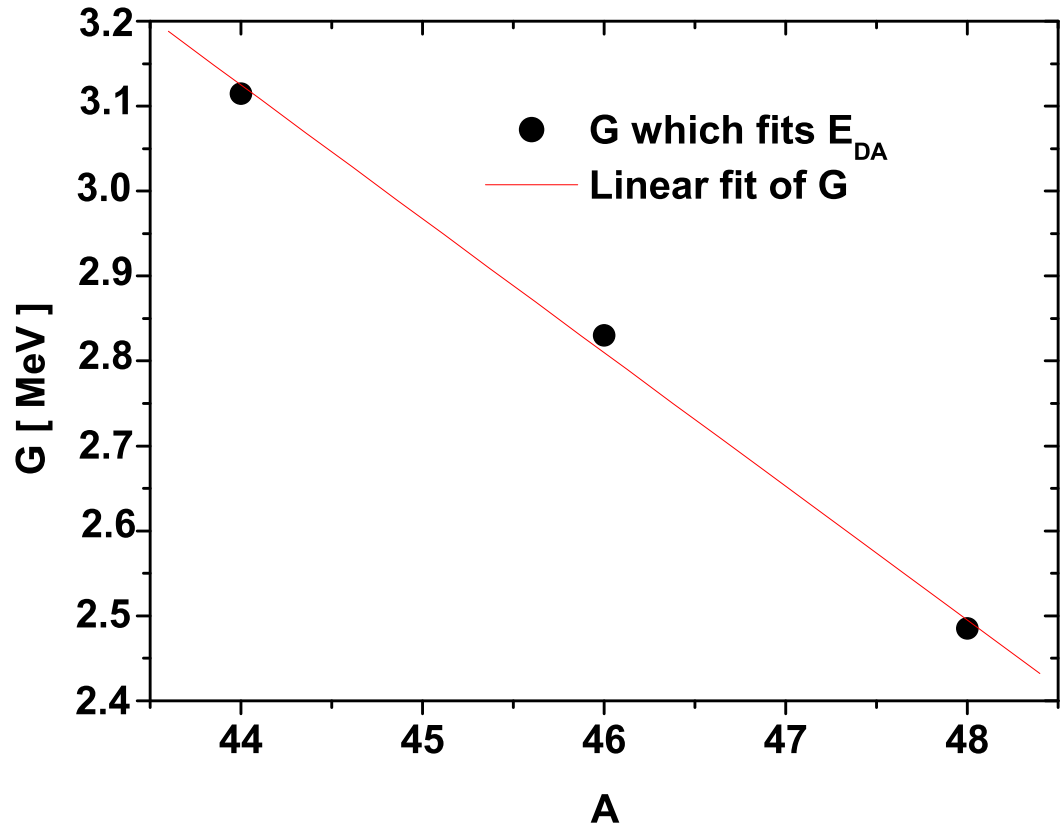, width=14cm, bbllx=2cm%
,bblly=5cm,bburx=20cm,bbury=26cm,angle=0}}
\vspace*{-3cm}
\caption{\label{Fig. 5}(Color online) The pairing strength as function of the
atomic mass A for even-mass isotopes of Ti.
The values of G are obtained by fitting the double analog resonance energies, respectively,
and represented by black circles. These points may be interpolated by a linear
polynomial in A. The resulting polynomial is given by the full line.}
\end{figure}

\newpage

\begin{figure}[h]
\centerline{\psfig{figure=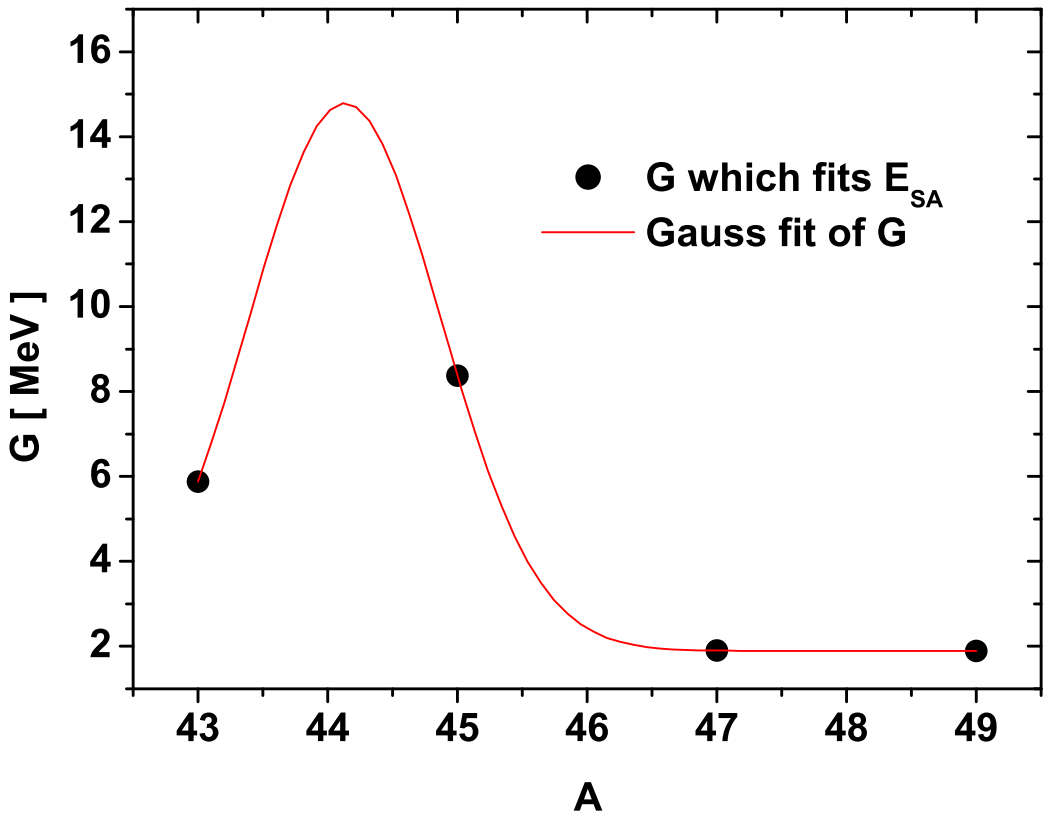, width=14cm, bbllx=2cm%
,bblly=5cm,bburx=20cm,bbury=26cm,angle=0}}
\vspace*{-3cm}
\caption{\label{Fig. 6} (Color online)The pairing strength as function of the
atomic mass A for odd-mass isotopes of Ti.
The values of G are obtained by fitting the single analog resonance energies, respectively,
and represented by black circles. These points may be interpolated by a Gauss
function of A. The resulting function is represented by a full line.}
\end{figure}

\newpage
\begin{figure}[h]
\vspace{2cm}
\begin{center}
\leavevmode
\epsfxsize = 12cm
\epsffile[30 4 518 361]{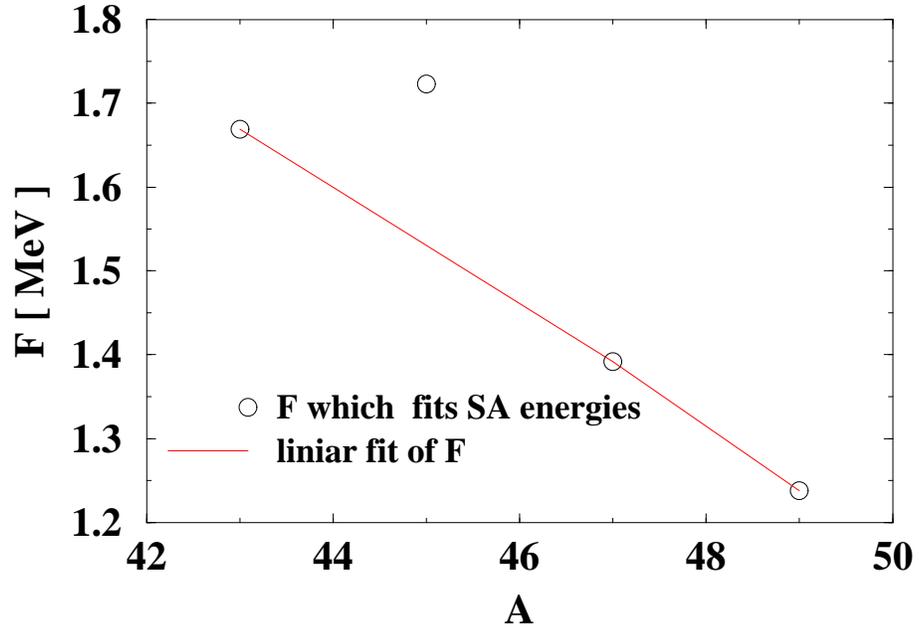}
\end{center}
\caption{(Color online)
The particle-core coupling strength as function of the
atomic mass A for odd-mass isotopes of Ti.
The values of F are obtained by fitting the single analog resonances energies, respectively, by using Eq. (18),
and represented by white circles. Three points lie on a straight line represented by a full line.
}
\label{fig7}
\end{figure}
\end{document}